\documentclass[twocolumn,showpacs,preprintnumbers,amsmath,amssymb,subfig]{revtex4}

\usepackage{graphicx}

\begin{document}

\title{Discontinuous classical ground state magnetic response as an even-odd effect in higher order rotationally invariant exchange interactions}

\author{N. P. Konstantinidis}
\affiliation{Department of Physics, King Fahd University of Petroleum and Minerals, Dhahran 31261, Saudi Arabia}

\date{\today}

\begin{abstract}
The classical ground state magnetic response of the Heisenberg model when rotationally invariant exchange interactions of integer order $q>1$ are added is found to be discontinuous, even though the interactions lack magnetic anisotropy. This holds even in the case of bipartite lattices which are not frustrated, as well as for the frustrated triangular lattice. The total number of discontinuities is associated with even-odd effects as it depends on the parity of $q$ via the relative strength of the bilinear and higher order exchange interactions, and increases with $q$. These results demonstrate that the precise form of the microscopic interactions is important for the ground state magnetization response.





\end{abstract}

\pacs{71.70.Gm Exchange Interactions, 75.10.Hk Classical Spin Models, 75.10.Pq Spin Chain Models}

\maketitle















The Heisenberg model plays an extremely important role in the study of magnetism of strongly correlated electron systems \cite{Auerbach98,Fazekas99,Lhuillier01,Misguich03}. Within its context discontinuities of the ground state magnetization were found, showing that it can be tuned between well-separated values with small changes of an external field. Such discontinuities naturally occur in the presence of magnetic anisotropy, where the field forces the spins to non-continuously change their orientation toward special directions in spin space along which the energy is more efficiently minimized \cite{Neel36}. Frustrated clusters are of special interest, as competing interactions can lead to discontinuous ground state magnetization response in the absence of anisotropy, with the ground state spin configuration completely changing its symmetry as the discontinuity is traversed \cite{Coffey92,NPK05,Schroder05,NPK07,NPK15,NPK16,NPK16-1,NPK17,NPK17-1}. A non-continuous magnetic response is therefore associated with preferential directions in spin space or non-trivial connectivity of the interacting spins.

Here a new source of discontinuities is identified, rotationally invariant exchange interactions of integer order $q>1$, which require neither magnetic anisotropy nor frustration to generate magnetization jumps. Such discontinuities occur for lattices as simple as the bipartite, as well as for the frustrated triangular lattice that was further considered. These interactions are associated with even-odd effects with respect to the parity of $q$: for positive higher order exchange they generate ground state discontinuities only when $q$ is odd, while for negative there are discontinuities for both even and odd $q$. In the latter case the total number of magnetization gaps also depends on the parity of $q$. For the same type of parity, the total number of discontinuities increases with $q$.

It has been shown that higher order exchange terms are often important for the explanation of experimental data \cite{Furrer11}. This has been the case with the biquadratic exchange interaction ($q=2$) \cite{Falk84,Gaulin86,Demokritov98,Millet99,Orzel01}, which can even be quite stronger than the bilinear exchange \cite{Nakatsuji05,Tsunetsugu06,Lauchli06-1}. Its importance can not be understated also from the theoretical point of view \cite{Anderson63}, for example for the chain hosting spins of magnitude $s=1$ \cite{Thorpe72,Haldane83,Haldane83-1,Affleck87,Lauchli06}, and it has also been considered along with biqubic terms for $s=\frac{3}{2}$ \cite{Fridman11}. It was also investigated in higher dimensions \cite{Hayden10,Kawamura07,Wenzel13}, and for the magnetism of iron-based supeconductors \cite{Yu15,Zhuo16}.
The classical ground state magnetic response with bilinear and biquadratic exchange has been calculated for short odd-numbered chains \cite{NPK15-1}, while in the case of the icosahedron it has been shown to generate many magnetization discontinuities \cite{NPK16-1}. A similar Hamiltonian in two dimensional spin space includes the standard bilinear exchange and generalized nematic interactions \cite{Lee85,Korshunov85,Park08,Poderoso11,Zukovic16}.

Here rotationally invariant exchange interactions with $q$ ranging from 2 to 9 are considered. For spins $s$ the highest order non-trivial exchange interaction term of this type is of order $2s$. This means that for higher $q$ the classical treatment of the problem where the bilinear competes with the $2s$-order exchange interaction provides a very good description of the quantum problem, and that the even-odd effects in $q$ can also be viewed as effects related to $s$ being integer or half-integer. The underlying lattices are of bipartite form, chains and rectangles, and the frustrated triangular lattice.

The Hamiltonian including bilinear and higher order exchange interactions as well as a magnetic field term is
\begin{equation}
H = \sum_{<ij>} [ J \vec{s}_i \cdot \vec{s}_j + J' ( \vec{s}_i \cdot \vec{s}_j )^q ] -  h \sum_{i=1}^{N} s_i^z
\label{eqn:model}
\end{equation}
There are $N$ spins $\vec{s}_i$ which are classical unit vectors in three-dimensional spin space. $<ij>$ indicates that interactions are limited to nearest neighbors $i$ and $j$. The first term is the bilinear exchange interaction, scaled with $J$, and the second the higher order exchange interaction of integer order $q>1$, scaled with $J'$. The exchange interactions are taken to be isotropic in spin space. The magnetic field $\vec{h}$ in the Zeeman term points along the $z$ direction without any loss of generality. The interactions are parametrized as $J$=cos$\omega$ and $J'$=sin$\omega$. The bilinear exchange favors antiparallel nearest-neighbors for positive $J$ and parallel for negative $J$. When $J'$ is positive the second term in Hamiltonian (\ref{eqn:model}) favors antiparallel nearest-neighbor spins for odd $q$ and perpendicular for even $q$. For negative $J'$ it favors ferromagnetically coupled spins irrespectively of the parity of $q$. The situation is further complicated by the Zeeman term, through which the spins gain maximum magnetic energy when pointing in the direction of the field. The competition of these three terms determines the magnetic properties. In addition, for lattices such as the triangular their frustrated connectivity plays an important role. A ground state magnetization discontinuity is associated with a non-continuous change of the lowest energy spin configuration, which originates in a more efficient energy minimization as the field increases. Here the lowest energy configuration of Hamiltonian (\ref{eqn:model}) is numerically calculated as a function of $h$ for $\omega$ $\epsilon$ $[0,2\pi)$ \cite{Coffey92,NPK05,NPK07,Machens13,NPK13,NPK15,NPK15-1,NPK16,NPK16-1,NPK17,NPK17-1}. The direction of each spin $\vec{s}_i$ is defined by a polar $\theta_i$ and an azimuthal $\phi_i$ angle. All angles are randomly initialized and each one is moved opposite its gradient direction until the energy minimum is reached. The procedure is repeated for different initial configurations to ensure that the global lowest energy configuration is found.

\begin{figure}[p]
\includegraphics[width=3.5in,height=2.5in]{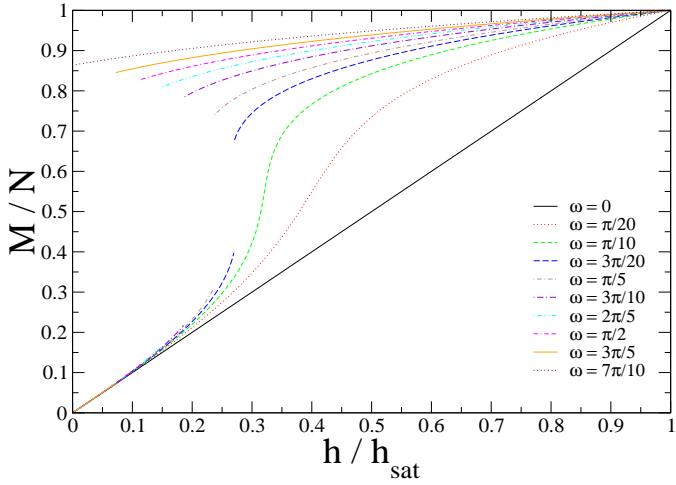}
\vspace{0pt}
\caption{(Color online) Magnetization per spin $\frac{M}{N}$ as a function of the magnetic field over its saturation value $\frac{h}{h_{sat}}$ in the lowest energy configuration of Hamiltonian (\ref{eqn:model}) for $q=3$ and different $\omega$ values for a chain or rectangle.
}
\label{fig:exchangeq=3}
\end{figure}

\begin{figure}[p]
\includegraphics[width=3.5in,height=2.5in]{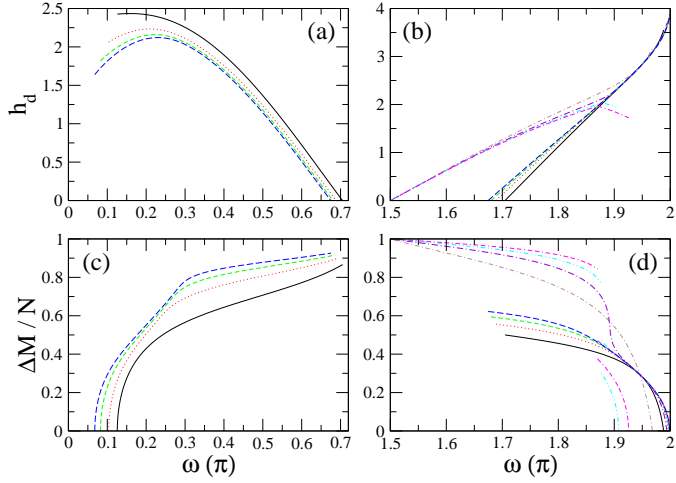}
\vspace{0pt}
\caption{(Color online) (a) Discontinuity magnetic fields $h_d$ in the lowest energy configuration of Hamiltonian (\ref{eqn:model}) for a chain or rectangle for lower $\omega$ and $q=3$ ((black) solid line), $q=5$ ((red) dotted line), $q=7$ ((green) dashed line), and $q=9$ ((blue) long-dashed line). (b) Similar with (a) for higher $\omega$ and additional lines for $q=2$ ((brown) dot-dashed line), $q=4$ ((violet) dot-long dashed line), $q=6$ ((cyan) double dot-dashed line), and $q=8$ ((magenta) dot-double dashed line). (c) Corresponding magnetization change per spin $\frac{\Delta M}{N}$ for (a). (d) Corresponding $\frac{\Delta M}{N}$ for (b).
}
\label{fig:qexchangewidthlocation}
\end{figure}

Firstly bipartite structures are considered, chains and rectangles with periodic boundary conditions, which lack any frustration.
Minimization of Hamiltonian (\ref{eqn:model}) shows that for lower and higher $\omega$, where the bilinear exchange is positive or weakly negative and the ground state is not ferromagnetic in zero field, the lowest energy configuration is the same for both types of structures for increasing $N$ and different $q$, and consequently also the one for the corresponding infinite lattices. The configuration is planar with a unit cell of two spins with the same polar angle $\theta$, and azimuthal angles that differ by $\pi$. Thus each spin's nearest-neighbors point in the same direction, while $\theta$ is given from the solution of the equation
\begin{eqnarray}
( 2 cos^2\theta -1)^{q-1} = \frac{1}{qJ'} ( \frac{h}{4 cos\theta} - J )
\label{eqn:eqntheta}
\end{eqnarray}
(see App. \ref{appendix:ZF-LEC-Bipartite} for the case of zero field). The solution of Eq. (\ref{eqn:eqntheta}) can be discontinuous at a magnetic field $h_d$. Fig. \ref{fig:exchangeq=3} plots the magnetization per spin $\frac{M}{N}=cos\theta$ as a function of $h$ for $q=3$. When $\frac{J'}{J}=0.41667$ (or $\omega=0.12567 \pi$) a discontinuity appears which originates in the higher order exchange interaction, as it is present when $J=0$ ($\omega=\frac{\pi}{2}$), and survives up to $\omega=0.70483\pi$.
Fig. \ref{fig:qexchangewidthlocation}(a) plots the discontinuity field for $J'>0$. The jump occurs for odd but not for even $q$, bringing about an even-odd effect in the parity of higher order exchange interactions. An odd $q$ favors antiparallel nearest-neighbors, while an even $q$ perpendicular. The discontinuity exists for a specific $\omega$ range (see also App. \ref{appendix:MagnetizationinaField-LEC-Bipartite}), and is always present for $J=0$ ($\omega=\frac{\pi}{2}$).
It requires neither magnetic anisotropy nor frustration in the interactions. A higher $q$ pushes it towards smaller magnetic fields and increases its width per spin $\frac{\Delta M}{N}$ (Fig. \ref{fig:qexchangewidthlocation}(c)).

For $J>0$ and $J'<0$ another magnetization discontinuity appears, irrespectively of the parity of $q$. The higher order exchange interactions favor parallel spins for any $q$ exactly like the magnetic field, and compete with the antiferromagnetic bilinear exchange. This competition is now the origin of the discontinuity. The lowest energy configuration changes abruptly from the one predicted by Eq. (\ref{eqn:eqntheta}) to the ferromagnetic one, with the corresponding discontinuity fields plotted in Fig. \ref{fig:qexchangewidthlocation}(b) (see also App. \ref{appendix:MagnetizationinaField-LEC-Bipartite}). There is another even-odd effect, with the discontinuity triggered by an infinitesimal positive $J$ for even $q$, while a finite $J$ value is needed for odd $q$. On the other hand only a small negative value of $J'$ is required to generate the jump close to the bilinear exchange limit for any $q$. Then only for even $q \geq 6$ and higher $\omega$ the discontinuity breaks up in two, with the one leading to saturation following the pattern of the odd $q$ and lower even $q$ discontinuities. This can also be seen in the plot of the width per spin of the jumps (Fig. \ref{fig:qexchangewidthlocation}(d)). These results show that the ground state magnetic response gets richer with $q$, demonstrating also the importance of the detailed form of the microscopic interactions and not only their symmetry for the precise determination of the magnetization curve \cite{Poderoso11}.

\begin{figure}[p]
\includegraphics[width=3.5in,height=2.5in]{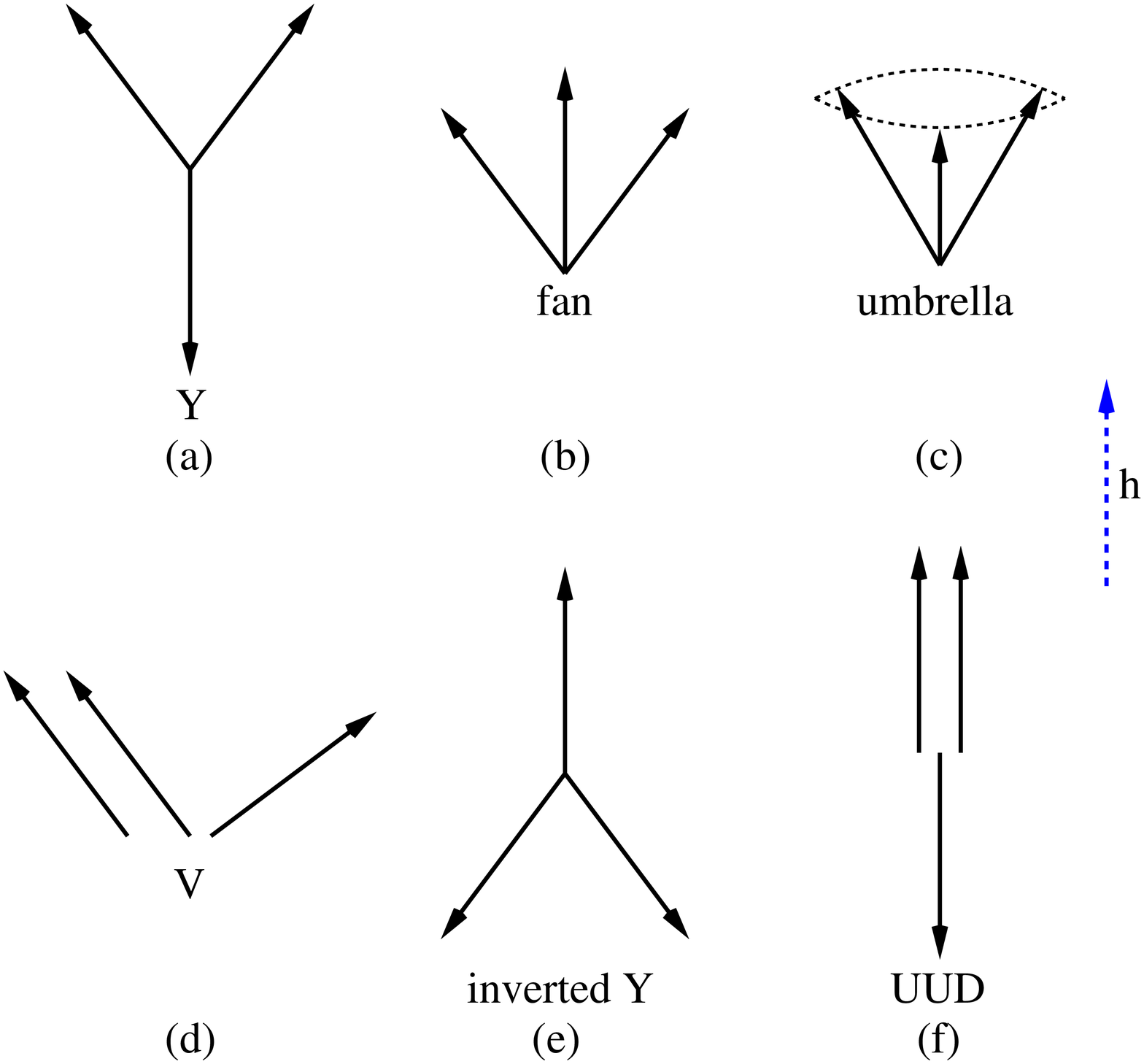}
\vspace{0pt}
\caption{(Color online) Lowest energy configurations of Hamiltonian (\ref{eqn:model}) for the triangular lattice.
}
\label{fig:qexchangetriangularlowestenergyconfigurations}
\end{figure}

\begin{figure}[p]
\begin{minipage}{\textwidth}
\centerline{
\includegraphics[width=0.3\textwidth,height=1.5in]{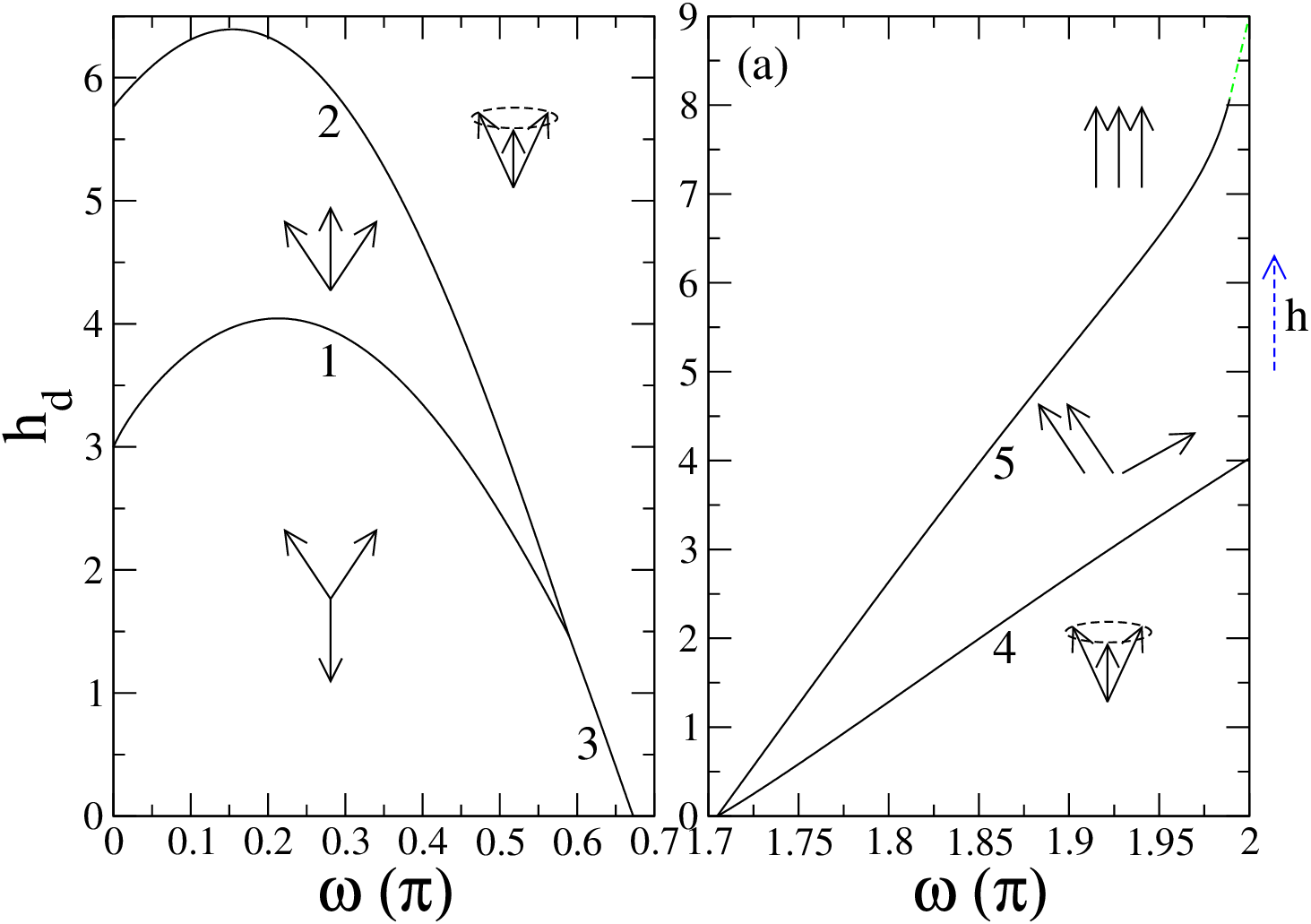}
\hspace{0pt}
\includegraphics[width=0.3\textwidth,height=1.5in]{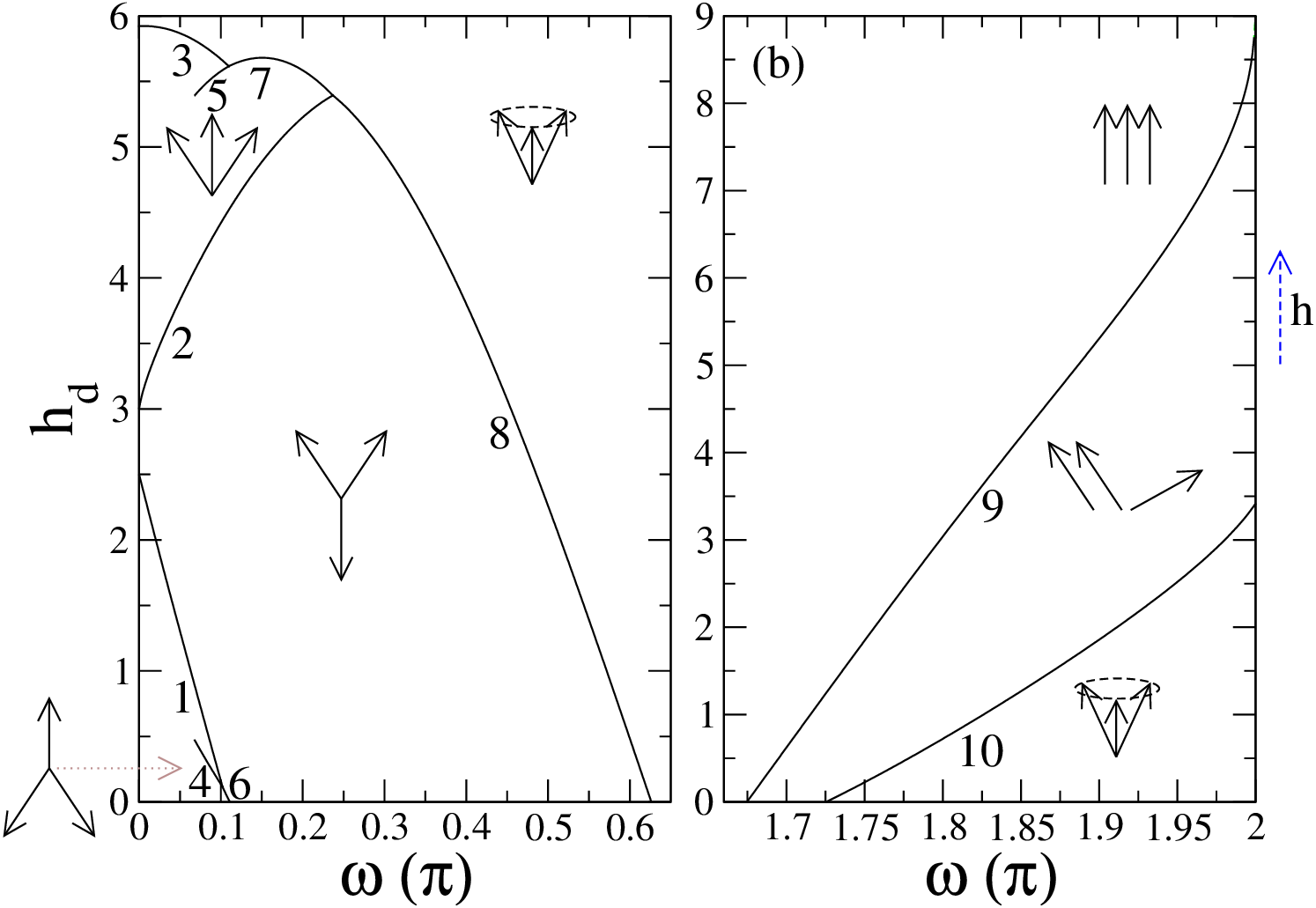}
}
\vspace{17pt}
\centerline{
\includegraphics[width=0.3\textwidth,height=1.5in]{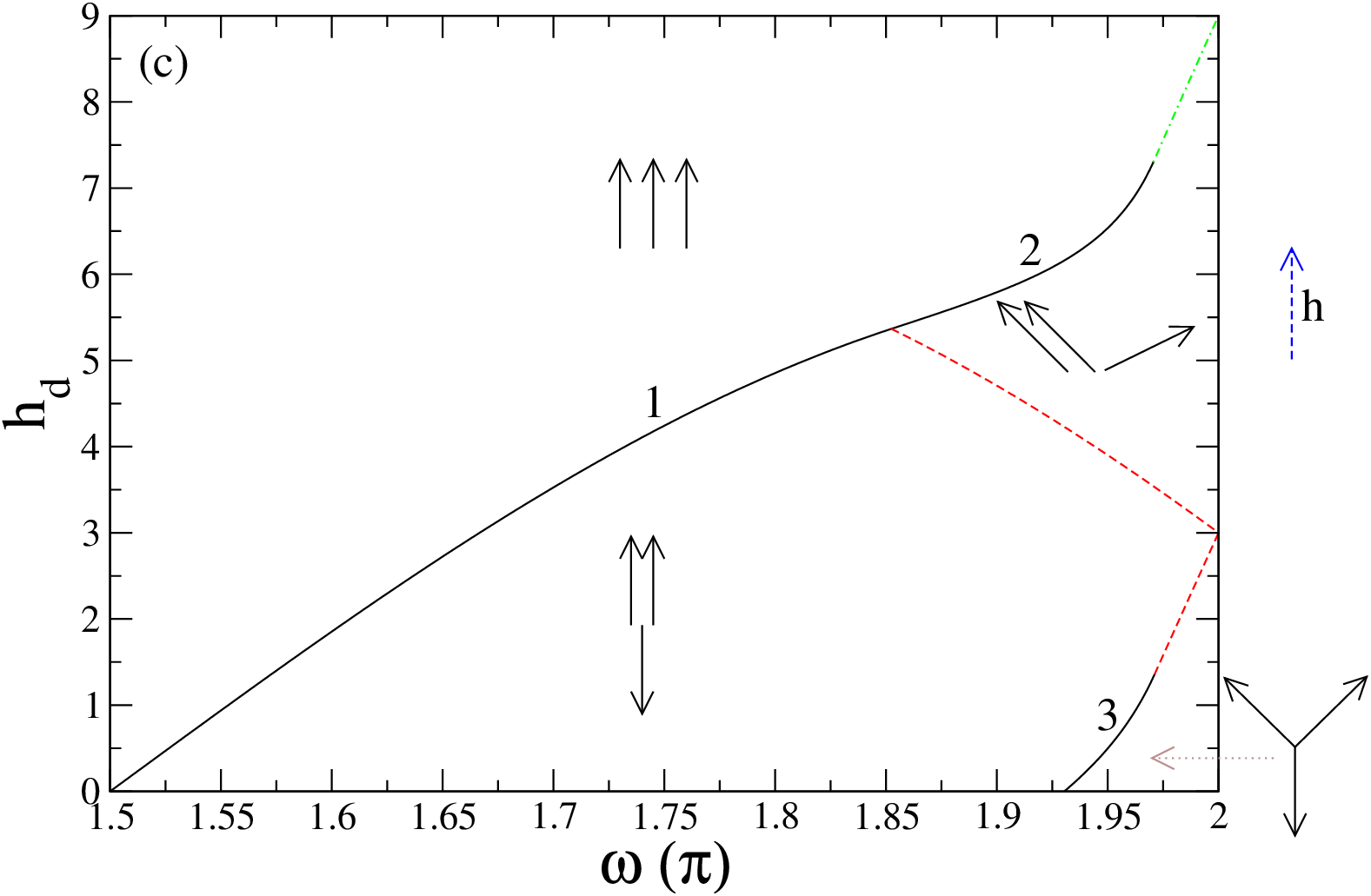}
\hspace{0pt}
\includegraphics[width=0.3\textwidth,height=1.5in]{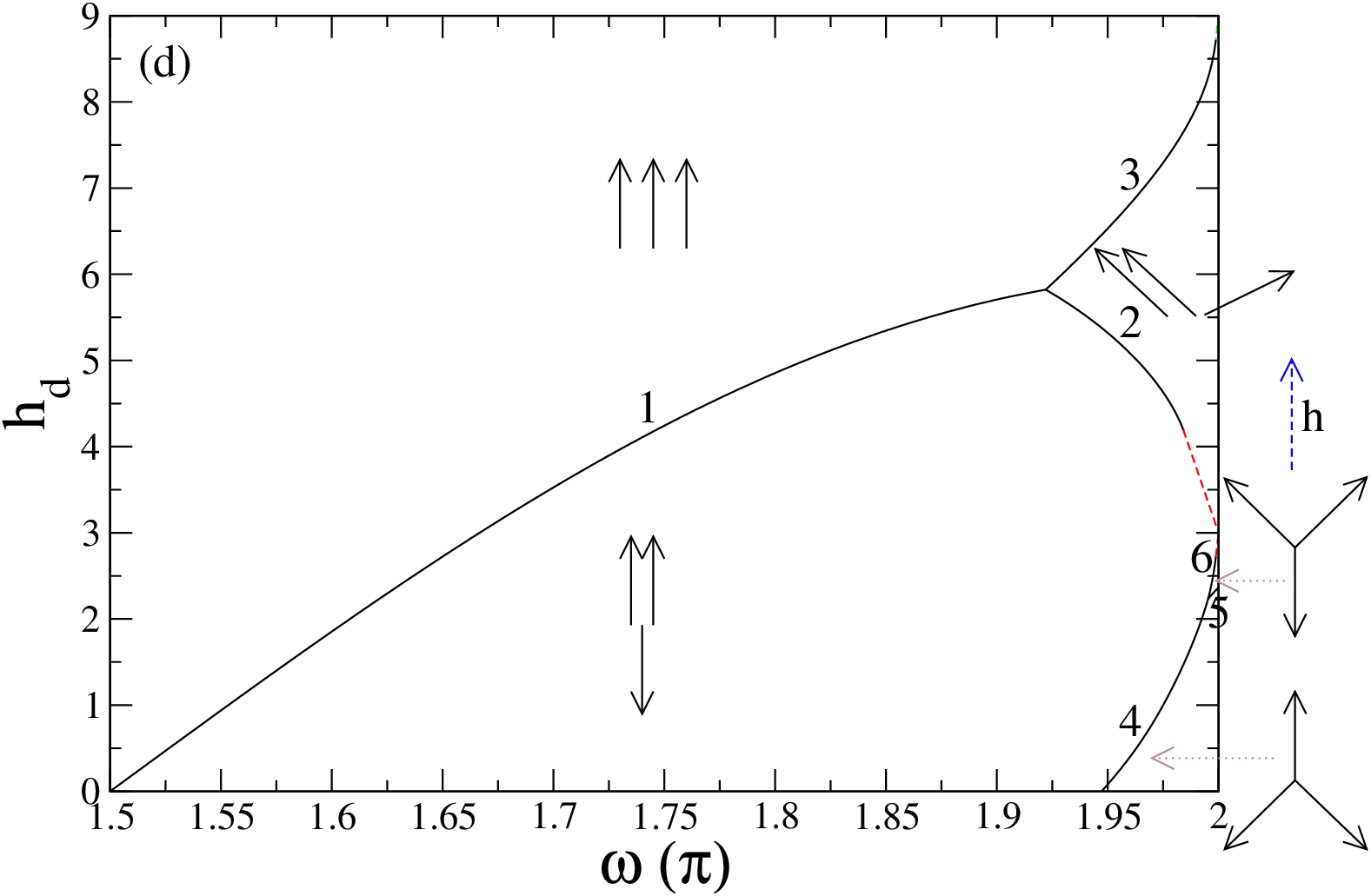}
}
\caption{(Color online) The (black) solid lines show the discontinuity magnetic fields $h_d$ in the lowest energy configuration of Hamiltonian (\ref{eqn:model}) for the triangular lattice as a function of $\omega$ for (a) $q=3$, (b) $q=9$, (c) $q=2$, and (d) $q=8$. The (red) dashed lines show the fields which are limits of the $\frac{1}{3}$ magnetization plateau. The (green) dot-dashed lines show the saturation field for the highest $\omega$. Each discontinuity is identified by a number.
}
\label{fig:qexchangetriangular}
\end{minipage}
\end{figure}

Hamiltonian (\ref{eqn:model}) is frustrated in the case of the triangular lattice. In the absence of higher order exchange interactions and in finite field it has an accidental classical ground state degeneracy that is lifted by thermal \cite{Kawamura84} and quantum fluctuations \cite{Chubukov91}. This order by disorder effect is also generated by nonmagnetic impurities \cite{Maryasin13} and anisotropic terms \cite{Griset11}. When higher order exchange interactions are included minimization of Hamiltonian (\ref{eqn:model}) with periodic boundary conditions shows that they also break the degeneracy and induce order (see App. \ref{appendix:ZF-LEC-Triangular} for the case of zero field). The ground state has a triangular unit cell, with spin configurations selected from the finite field ground state degenerate manifold of $\omega=0$ and plotted in Fig. \ref{fig:qexchangetriangularlowestenergyconfigurations} (see App. \ref{appendix:MagnetizationinaField-LEC-Triangular}). The frustrated connectivity of the triangular lattice generates a ground state magnetization response with more discontinuities than the one of the bipartite lattices. Such discontinuities were first found for a finite version of the triangular lattice, the icosahedron, already for $q=2$ \cite{NPK16-1}. Again there is an even-odd effect for $J'>0$, with discontinuities occuring only for odd $q$. Figure \ref{fig:qexchangetriangular}(a) plots the corresponding fields for $q=3$. An infinitesimal positive $J'$ is sufficient to generate two magnetization jumps, which merge for higher $\omega$. The lower field discontinuity changes the configuration from the Y to the fan,
while the higher field jump leads to the non-coplanar
''umbrella'' configuration. The higher order exchange interaction generates the discontinuities, with both occuring when the bilinear exchange $J=0$ ($\omega=\frac{\pi}{2}$). This time the frustrated connectivity allows the jumps to survive down to infinitesimal $J'$. When $J>0$ and $J'<0$
two discontinuities occur. The lower changes the spin configuration from the ''umbrella'' to the V. This magnetization gap also requires an infinitesimal value of (negative) $J'$ to occur. The higher field jump leads the spins directly to the ferromagnetic configuration, similarly to the case of the bipartite lattices. A jump to saturation has been found in the quantum case for the antiferromagnetic Heisenberg model in frustrated lattices and molecules \cite{Schulenburg02,Schnack01,Schmidt05}. It is stressed that infinitesimal deviations from the purely bilinear exchange case generate discontinuous ground state response irrespectively of the sign of $J'$.

Figure \ref{fig:qexchangetriangular}(b) plots the discontinuity fields when $q=9$. Contrary to the case of the bipartite lattices the discontinuity diagram becomes richer for $J'>0$, with a maximum of five discontinuities for $0.067898 \pi \leq \omega \leq 0.10199 \pi$. An infinitesimal $J'$ generates now a total of three jumps, with the inverted Y configuration appearing for small fields. The jumps are now related not only to a change of the configuration type but also to discontinuous polar angles within the same type of configuration (discontinuities 4 and 5). These results show as in the case of the bipartite lattices that the precise value of $q$ is important for the determination of the ground state magnetic response.

\begin{figure}
\begin{minipage}{\textwidth}
\centerline{
\includegraphics[width=0.3\textwidth,height=1.5in]{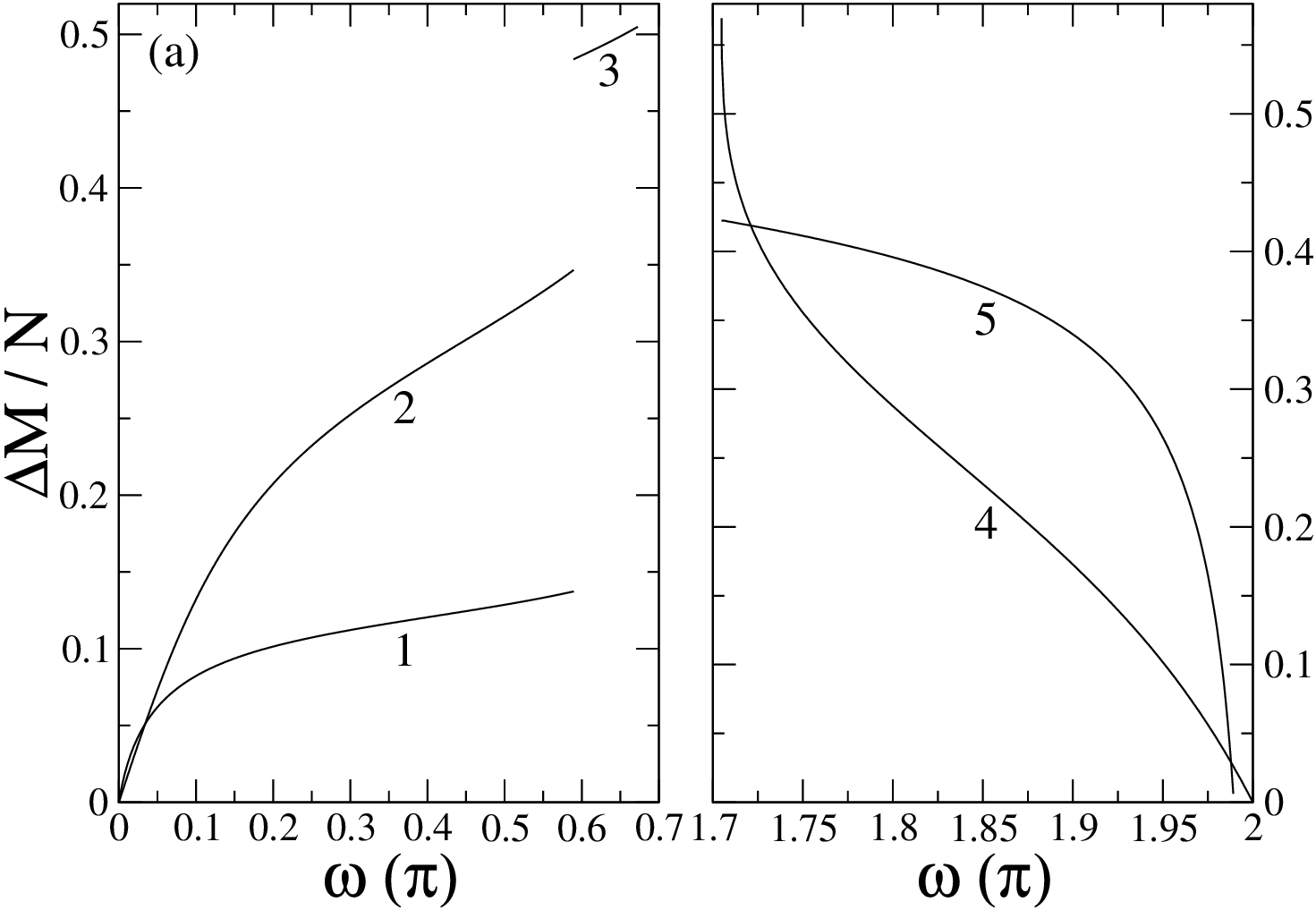}
\hspace{0pt}
\includegraphics[width=0.3\textwidth,height=1.5in]{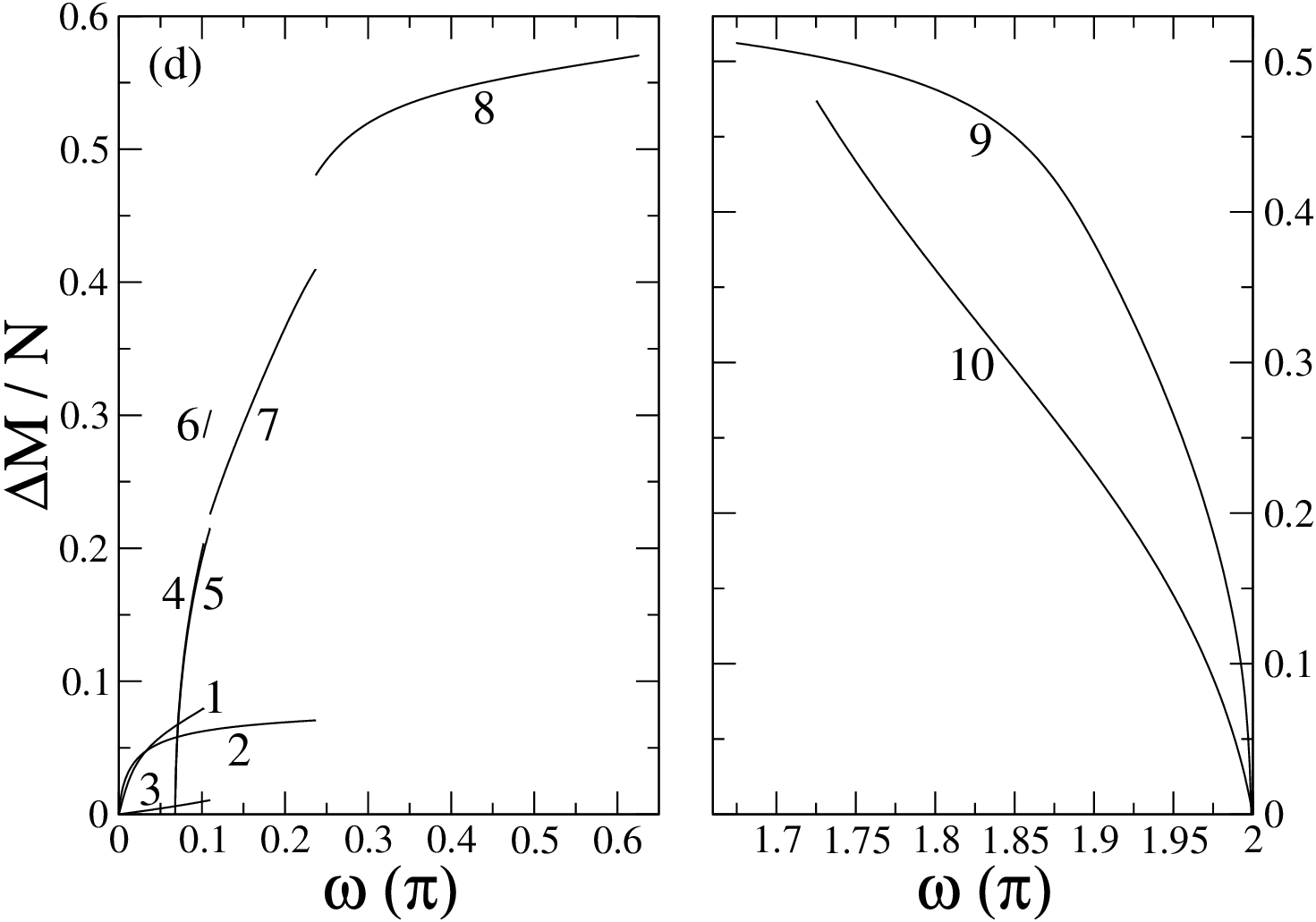}
}
\vspace{17pt}
\centerline{
\includegraphics[width=0.3\textwidth,height=1.5in]{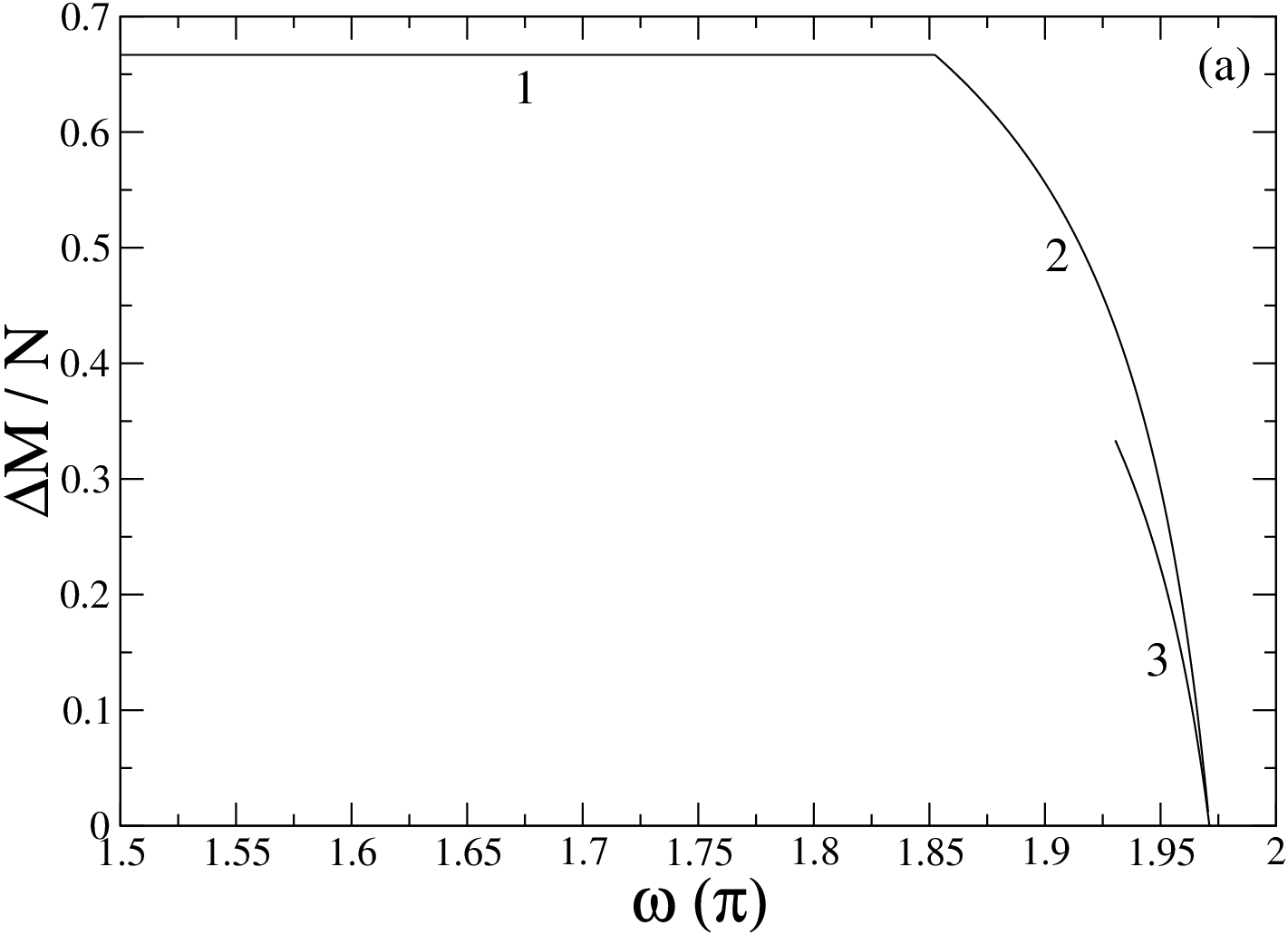}
\hspace{0pt}
\includegraphics[width=0.3\textwidth,height=1.5in]{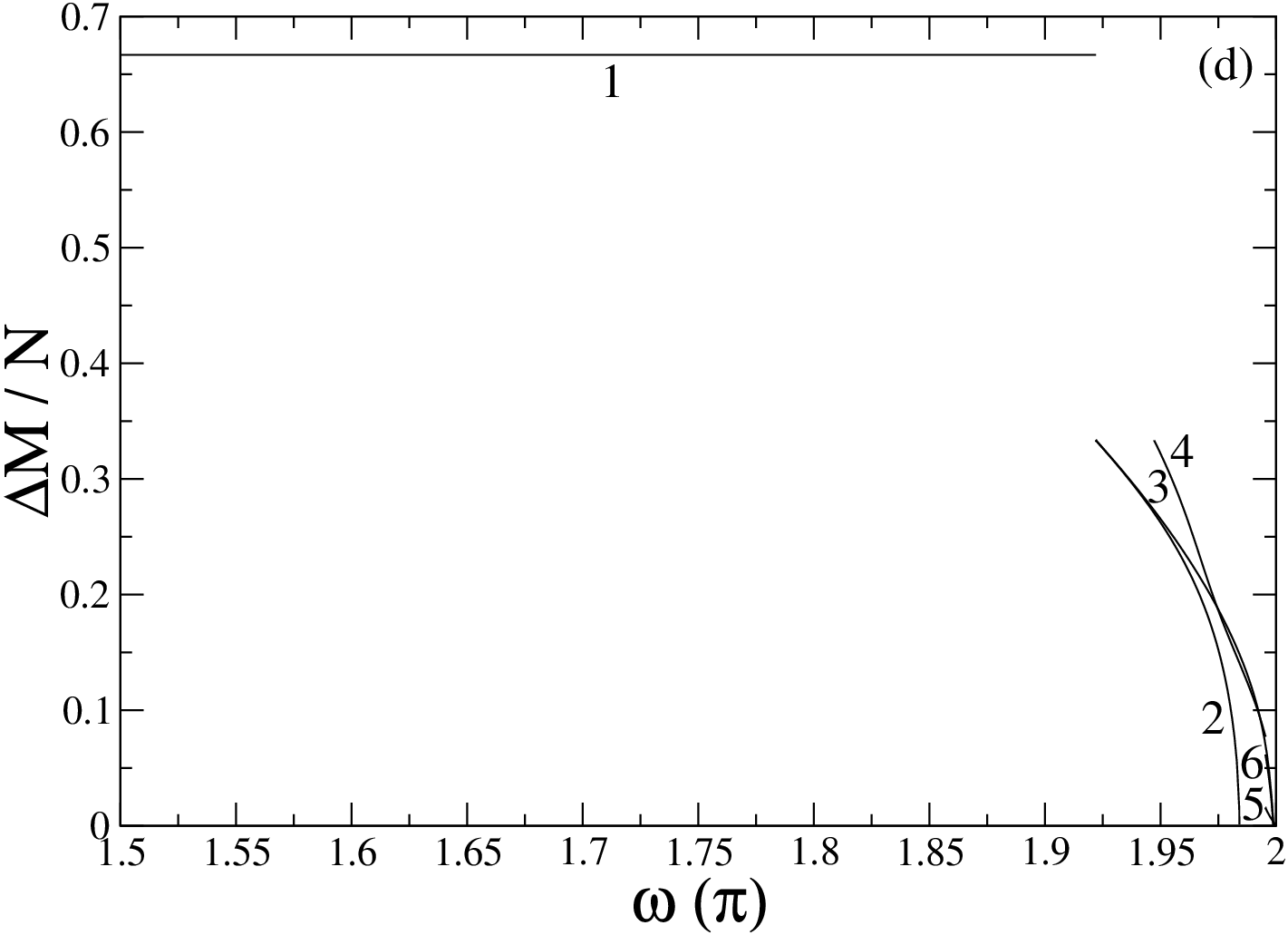}
}
\caption{Magnetization change per spin $\frac{\Delta M}{N}$ for the discontinuities in the lowest energy configuration of Hamiltonian (\ref{eqn:model}) for the triangular lattice as a function of $\omega$ for (a) $q=3$, (b) $q=9$, (c) $q=2$, and (d) $q=8$. The numbering of the discontinuities follows Figs \ref{fig:qexchangetriangular}(a), \ref{fig:qexchangetriangular}(b), \ref{fig:qexchangetriangular}(c), and \ref{fig:qexchangetriangular}(d).
}
\label{fig:qexchangetriangular2}
\end{minipage}
\end{figure}

Figure \ref{fig:qexchangetriangular}(c) shows the discontinuities for $q=2$. An infinitesimal deviation from $\omega=\frac{3\pi}{2}$ generates a jump by changing the configuration from the UUD associated with the $\frac{1}{3}$ magnetization plateau of the triangular lattice to the saturated one. The highest field discontinuities direct all spins to be parallel to the field, as in the bipartite lattices case. An infinitesimal (negative) $J'$ generates the $\frac{1}{3}$ magnetization plateau as indicated by the (red) dashed lines. This plateau is a feature of the antiferromagnetic Heisenberg model in the quantum case \cite{Schnack01,Honecker04}. For small negative $J'$ the field drives the system from the Y to the UUD and then to the V configuration, similarly to the effect of finite temperature in the $J'=0$ case \cite{Kawamura84}. In Fig. \ref{fig:qexchangetriangular}(d) it is shown that a higher value of $q=8$ again enriches the magnetization response generating more jumps, with the inverted Y configuration entering again for small fields and higher $\omega$. Again the response for $J>0$ and $J'<0$ showcases an even-odd effect with respect to $q$. The discontinuity strengths corresponding to Fig. \ref{fig:qexchangetriangular} are plotted in Fig. \ref{fig:qexchangetriangular2} (for the rest of the $q$ values see App. \ref{appendix:MagnetizationinaField-LEC-Triangular}).

In conclusion, the classical ground state magnetization response has been calculated for bipartite lattices and the triangular lattice, when isotropic exchange interactions of integer order $q>1$ compete with the standard bilinear exchange interaction. These interactions generate magnetization discontinuities even though there is no magnetic anisotropy or necessarily frustration. The total number of discontinuities is associated with even-odd effects in $q$, and also increases with $q$. These results indicate that the precise form of the interactions and not only the symmetry of the Hamiltonian is important for the determination of zero-temperature properties, especially since a general interaction between spins can be expressed as a series expansion in powers of $q$. This is also expected for the thermodynamic properties where all the states are involved, as was shown in the case of nematic interactions \cite{Poderoso11}. Similar calculations can be performed for frustrated clusters where the addition of biquadratic exchange along with the special connectivity has led to a multitude of discontinuities already for a cluster as small as the icosahedron \cite{NPK16-1}, and more specifically for an even $q$ and bilinear and biquadratic exchange both positive, something not possible for the bipartite and triangular lattices.


\bibliography{qexchange}


\begin{appendix}

\section{Zero Field Lowest Energy Configuration for Bipartite Lattices}
\label{appendix:ZF-LEC-Bipartite}

The lowest energy configuration in the absence of a field is given by the solution of Eq. (\ref{eqn:eqntheta}) for $h=0$. The range of $\omega$ for which the lowest energy configuration is antiferromagnetic (AFM) and ferromagnetic (FM) for $2 \leq q \leq 9$ for a bipartite lattice is listed in Table \ref{table:bipartitezerofield}.

\begin{table}[p]
\begin{center}
\caption{Order of exchange interaction $q$ in Hamiltonian (\ref{eqn:model}) and corresponding ranges of $\omega$ for an AFM and a FM lowest energy state in zero magnetic field for a bipartite lattice. For odd $q$ the values on the left-hand sides equal $tan^{-1}(-\frac{1}{q})$. $0.70483\pi$ and $1.70483\pi$ are equal to $tan^{-1}(-\frac{4}{3})$.}
\begin{tabular}{c|c|c}
$q$ & AFM & FM \\
\hline
2 & $ \frac{3}{2} \pi < \omega \leq 0.14758 \pi$ & $0.85242 \pi \leq \omega \leq \frac{3}{2} \pi$ \\
\hline
3 & $1.89758 \pi < \omega \leq 0.70483\pi$ & $0.89758 \pi < \omega \leq 1.70483\pi$ \\
\hline
4 & $\frac{3}{2} \pi < \omega \leq 0.077979 \pi$ & $0.92202 \pi < \omega \leq \frac{3}{2} \pi$ \\
\hline
5 & $1.93717 \pi \leq \omega < 0.68868\pi$ & $0.93717 \pi \leq \omega < 1.68868\pi$ \\
\hline
6 & $\frac{3}{2} \pi < \omega \leq 0.052568 \pi$ & $0.94743 \pi < \omega \leq \frac{3}{2} \pi$ \\
\hline
7 & $1.95483 \pi < \omega < 0.68011 \pi$ & $0.95483 \pi < \omega < 1.68011 \pi$ \\
\hline
8 & $\frac{3}{2} \pi < \omega \leq 0.039583 \pi$ & $0.96042 \pi \leq \omega \leq \frac{3}{2} \pi$ \\
\hline
9 & $1.96478 \pi \leq \omega \leq 0.67471 \pi$ & $0.96478 \pi \leq \omega \leq 1.67471 \pi$ \\
\end{tabular}
\label{table:bipartitezerofield}
\end{center}
\end{table}

\section{Lowest Energy Configuration Magnetization in a Field for Bipartite Lattices}
\label{appendix:MagnetizationinaField-LEC-Bipartite}

Table \ref{table:bipartiteoddqdiscontinuities} lists the $\omega$ range of the two discontinuities for odd $q$ for a bipartite lattice, while Table \ref{table:bipartiteevenqdiscontinuities} lists the $\omega$ range of the discontinuities for even $q$. The saturation field $h_{sat}=4(J+qJ')$ except when the zero field ground state is ferromagnetic or a discontinuity leads directly to saturation.

\begin{table}[p]
\begin{center}
\caption{Odd order of exchange interaction $q$ in Hamiltonian (\ref{eqn:model}) and corresponding ranges of $\omega$ for the two magnetization discontinuities for a bipartite lattice. $0.70483\pi$ and $1.70483\pi$ are equal to $tan^{-1}(-\frac{4}{3})$.}
\begin{tabular}{c|c|c}
$q$ & disc. 1 & disc. 2 \\
\hline
3 & $0.12567 \pi < \omega \leq 0.70483 \pi$ & $1.70483\pi < \omega < 1.98822 \pi$ \\
\hline
5 & $0.10361 \pi \leq \omega < 0.68868\pi$ & $1.68868\pi \leq \omega < 1.99626 \pi$ \\
\hline
7 & $0.082519 \pi \leq \omega < 0.68011 \pi$ & $1.68011 \pi \leq \omega \leq 1.99818 \pi$ \\
\hline
9 & $0.067793 \pi < \omega \leq 0.67471 \pi$ & $1.67471 \pi < \omega < 1.99893 \pi$ \\
\end{tabular}
\label{table:bipartiteoddqdiscontinuities}
\end{center}
\end{table}


\begin{table}[p]
\begin{center}
\caption{Even order of exchange interaction $q$ in Hamiltonian (\ref{eqn:model}) and corresponding ranges of $\omega$ for the magnetization discontinuities for a bipartite lattice.}
\begin{tabular}{c|c|c|c}
$q$ & disc. 1 & disc. 2 & disc. 3 \\
\hline
2 & $\frac{3}{2} \pi < \omega \leq 1.96827 \pi$ & - & - \\
\hline
4 & $\frac{3}{2} \pi < \omega < 1.99388 \pi$ & - & - \\
\hline
6 & $\frac{3}{2} \pi < \omega < 1.87724 \pi$ & $1.87724 \pi \leq \omega < 1.90785 \pi$ & $1.87724 \pi \leq \omega \leq 1.99747 \pi$ \\
\hline
8 & $\frac{3}{2} \pi < \omega \leq 1.87042 \pi$ & $1.87042 \pi < \omega < 1.92550 \pi$ & $1.87042 \pi < \omega < 1.99863 \pi$ \\
\end{tabular}
\label{table:bipartiteevenqdiscontinuities}
\end{center}
\end{table}

\section{Zero Field Lowest Energy Configuration for the Triangular Lattice}
\label{appendix:ZF-LEC-Triangular}

The range of $\omega$ for which the lowest energy configuration in zero magnetic field has spins at $120^o$ degrees with each other, is FM or of the UUD form for $2 \leq q \leq 9$ for the triangular lattice is listed in Tables \ref{table:triangularzerofieldeven} and \ref{table:triangularzerofieldodd} (results for $q=2$ have been presented in Ref. \cite{Kawamura07}).


\begin{table}[p]
\begin{center}
\caption{Ranges of $\omega$ for an $120^0$, a FM and an UUD lowest energy state in zero magnetic field of Hamiltonian (\ref{eqn:model}) for even $q$ for the triangular lattice.}
\begin{tabular}{c|c|c|c}
$q$ & $120^o$ & FM & UUD \\
\hline
2 & $ 1.93040 \pi \leq \omega \leq \frac{\pi}{4}$ & $0.85242 \pi < \omega \leq \frac{3}{2} \pi$ & $\frac{3}{2} \pi < \omega < 1.93040 \pi$ \\
\hline
4 & $1.94400 \pi \leq \omega < 0.35242\pi$ & $0.92202\pi < \omega \leq \frac{3}{2} \pi$ & $\frac{3}{2} \pi < \omega < 1.94400 \pi$ \\
\hline
6 & $1.94661 \pi < \omega \leq 0.44100 \pi$ & $0.94743 \pi < \omega \leq \frac{3}{2} \pi$ & $\frac{3}{2} \pi < \omega \leq 1.94661 \pi$ \\
\hline
8 & $1.94723 \pi \leq \omega \leq 0.48013 \pi$ & $0.96042 \pi \leq \omega \leq \frac{3}{2} \pi$ & $\frac{3}{2} \pi < \omega < 1.94723 \pi$ \\
\end{tabular}
\label{table:triangularzerofieldeven}
\end{center}
\end{table}


\begin{table}[p]
\begin{center}
\caption{Ranges of $\omega$ for an $120^0$ and a FM lowest energy state in zero magnetic field of Hamiltonian (\ref{eqn:model}) for odd $q$ for the triangular lattice. For odd $q$ the values on the left-hand sides of the FM state equal $tan^{-1}(-\frac{1}{q})$. $1.70483\pi$ is equal to $tan^{-1}(-\frac{4}{3})$.}
\begin{tabular}{c|c|c}
$q$ & $120^o$ & FM \\
\hline
3 & $1.70483 \pi < \omega < 0.078728 \pi$ & $0.89758 \pi < \omega \leq 1.70483 \pi$ \\
\hline
5 & $1.69919 \pi < \omega < 0.085952 \pi$ & $0.93717 \pi \leq \omega < 1.68868 \pi$ \\
\hline
7 & $1.71103 \pi < \omega \leq 0.10312 \pi$ & $0.95483 \pi < \omega < 1.68011 \pi$ \\
\hline
9 & $1.72534 \pi \leq \omega < 0.11086 \pi$ & $0.96478 \pi \leq \omega \leq 1.67471 \pi$ \\
\end{tabular}
\label{table:triangularzerofieldodd}
\end{center}
\end{table}

\section{Lowest Energy Configuration Magnetization in a Field for the Triangular Lattice}
\label{appendix:MagnetizationinaField-LEC-Triangular}

\begin{figure}
\begin{minipage}{\textwidth}
\centerline{
\includegraphics[width=0.3\textwidth,height=1.5in]{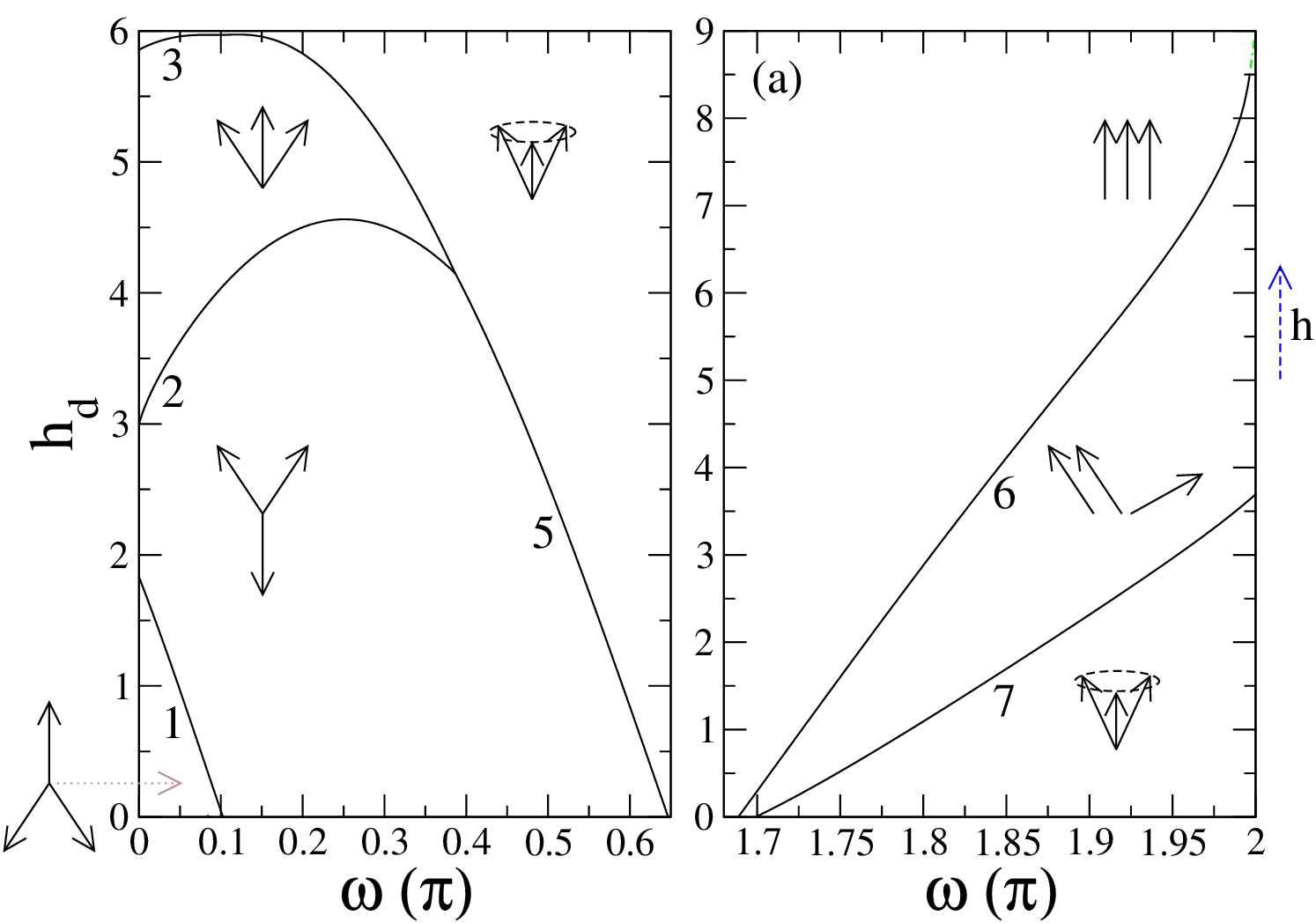}
\hspace{0pt}
\includegraphics[width=0.3\textwidth,height=1.5in]{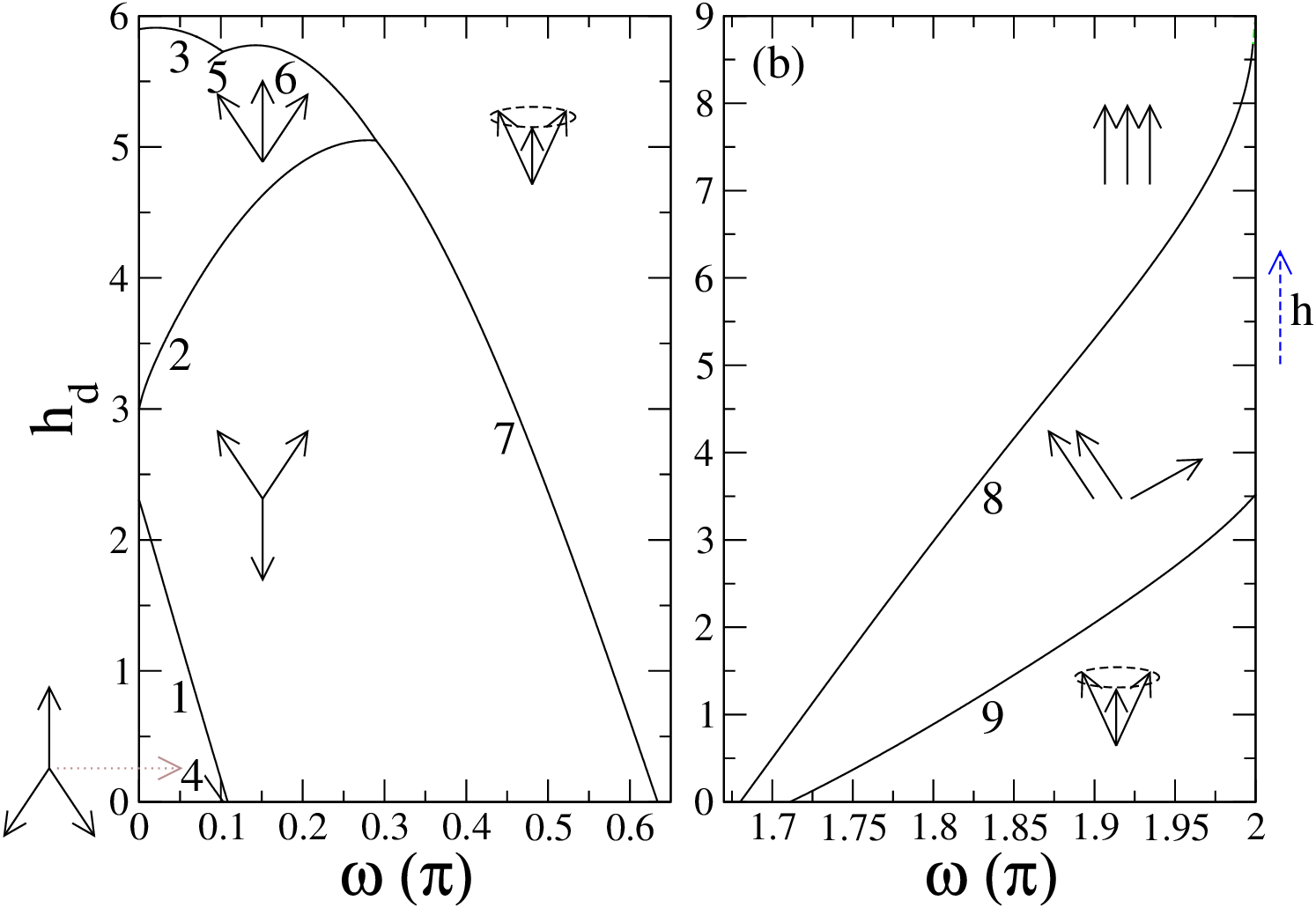}
}
\vspace{17pt}
\centerline{
\includegraphics[width=0.3\textwidth,height=1.5in]{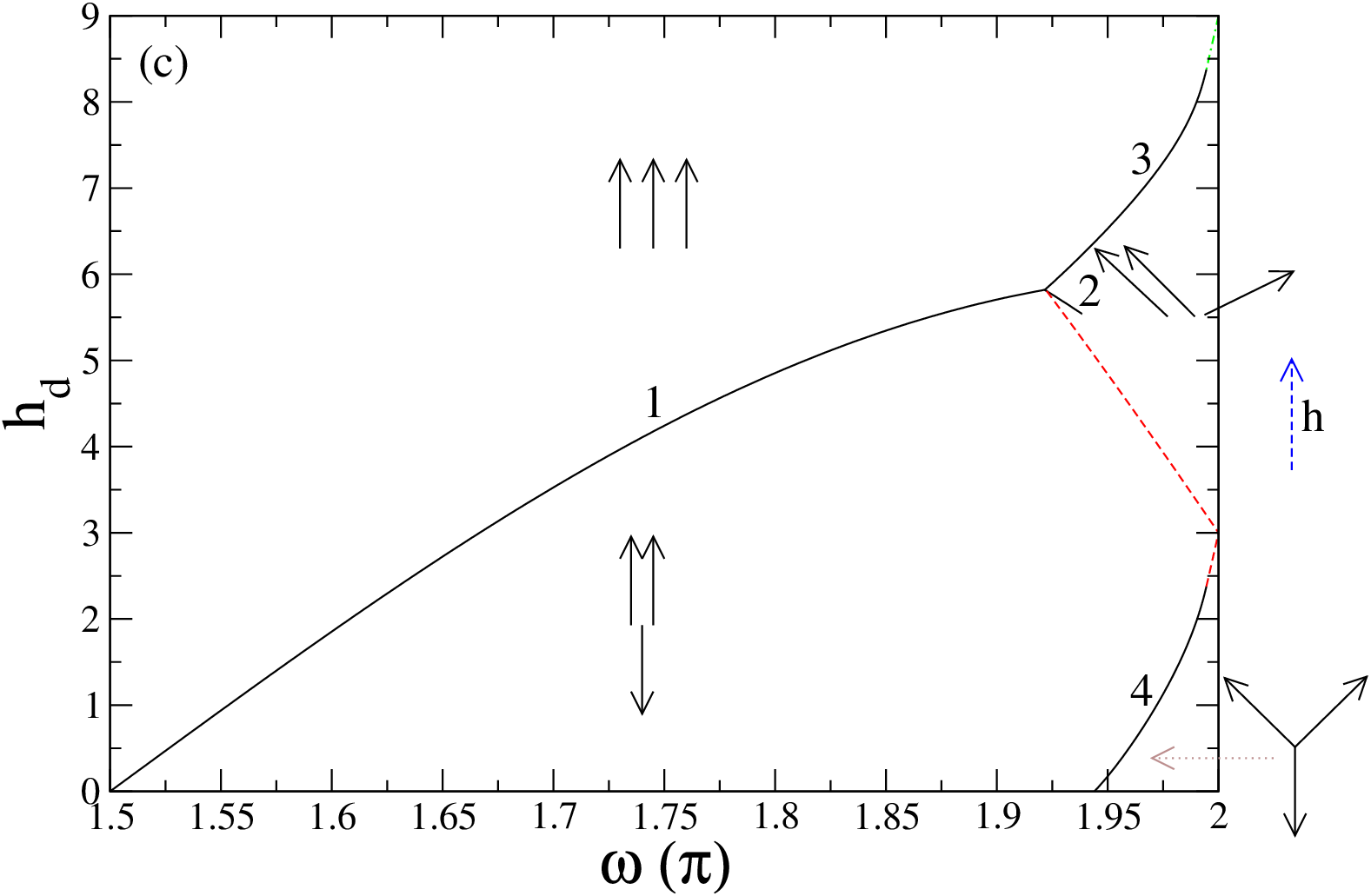}
\hspace{0pt}
\includegraphics[width=0.3\textwidth,height=1.5in]{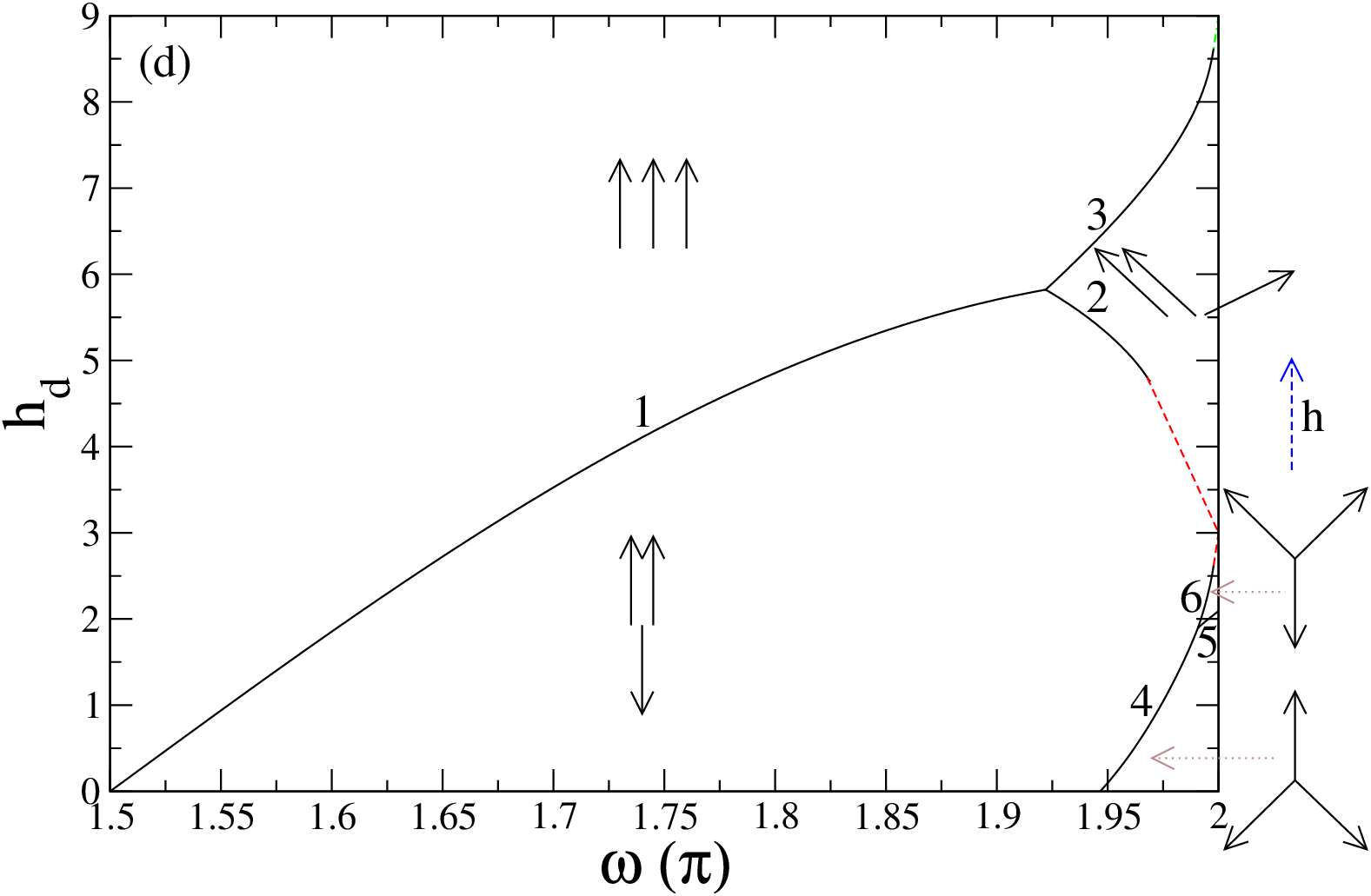}
}
\caption{(Color online) The (black) solid lines show the discontinuity magnetic fields $h_d$  in the lowest energy configuration of Hamiltonian (\ref{eqn:model}) for the triangular lattice as a function of $\omega$ for (a) $q=5$, (b) $q=7$, (c) $q=4$, and (d) $q=6$. The (red) dashed lines show the fields which are the limits of the $\frac{1}{3}$ magnetization plateau. The (green) dot-dashed lines show the saturation field for the highest $\omega$. Each discontinuity is identified by a number.
}
\label{fig:qexchangetriangular1}
\end{minipage}
\end{figure}

Table \ref{table:triangularoddqdiscontinuities} lists the $\omega$ range of the discontinuities for odd $q$ for the triangular lattice, while Table \ref{table:triangularevenqdiscontinuities} lists the $\omega$ range of the discontinuities for even $q$. Table \ref{table:triangularevenqplateaus} lists the ranges of $\omega$ for the limits of the magnetization plateaus for the triangular lattice. The saturation field $h_{sat}=9(J+qJ')$ except when the zero field ground state is ferromagnetic or a discontinuity leads directly to saturation. Figure \ref{fig:qexchangetriangular1} shows the discontinuity magnetic fields for $q=4$, 5, 6, and 7, and Fig. \ref{fig:qexchangetriangular3} the corresponding magnetization changes per spin. Fig. \ref{fig:qexchangetriangular4} shows one of the $q=5$ discontinuties in greater detail.

\begin{table}[p]
\begin{minipage}{\textwidth}
\begin{center}
\caption{Exchange interaction of odd order $q$ in Hamiltonian (\ref{eqn:model}) and corresponding ranges of $\omega$ for the magnetization discontinuities for the triangular lattice.}
\begin{tabular}{c|c|c|c|c}
disc. & $q=3$ & $q=5$ & $q=7$ & $q=9$ \\
\hline
1 & $0 < \omega \leq 0.58941 \pi$ & $0 < \omega \leq 0.10255 \pi$ & $0 < \omega \leq 0.10838 \pi$ & $0 < \omega \leq 0.10199 \pi$ \\
\hline
2 & $0 < \omega \leq 0.58941 \pi$ & $0 < \omega \leq 0.38655 \pi$ & $0 < \omega < 0.28976 \pi$ & $0 < \omega \leq 0.23687 \pi$ \\
\hline
3 & $ 0.58941 \pi < \omega < 0.67259 \pi$ & $0 < \omega \leq 0.38655 \pi$ & $0 < \omega < 0.10257 \pi$ & $0 < \omega < 0.10968 \pi$ \\
\hline
4 & $1.70483 \pi < \omega < 2\pi$ & $0.083764 \pi \leq \omega < 0.085952 \pi$ & $0.080520 \pi \leq \omega \leq 0.10312 \pi$ & $0.067689 \pi \leq \omega \leq 0.10199 \pi$ \\
\hline
5 & $1.70483 \pi < \omega \leq 1.98909 \pi$ & $0.38655 \pi < \omega < 0.64681 \pi$ & $0.084496 \pi \leq \omega < 0.10257 \pi$ & $0.067898 \pi \leq \omega < 0.10968 \pi$ \\
\hline
6 & & $1.68868 \pi \leq \omega \leq 1.99646 \pi$ & $0.10257 \pi \leq \omega < 0.28976 \pi$ & $0.10199 \pi < \omega < 0.11086 \pi$ \\
\hline
7 & & $1.69919 \pi < \omega < 2\pi$ & $0.28976 \pi \leq \omega < 0.63397 \pi$ & $0.10968 \pi \leq \omega \leq 0.23687 \pi$ \\
\hline
8 & & & $1.68011 \pi \leq \omega < 1.99838 \pi$ & $0.23687 \pi < \omega < 0.62624 \pi$ \\
\hline
9 & & & $1.71103 \pi < \omega < 2\pi$ & $1.67471 \pi < \omega \leq 1.99904 \pi$ \\
\hline
10 & & & & $1.72534 \pi \leq \omega < 2\pi$ \\
\end{tabular}
\label{table:triangularoddqdiscontinuities}
\end{center}
\end{minipage}
\end{table}

\begin{table}[p]
\begin{minipage}{\textwidth}
\begin{center}
\caption{Exchange interaction of even order $q$ in Hamiltonian (\ref{eqn:model}) and corresponding ranges of $\omega$ for the magnetization discontinuities for the triangular lattice.}
\begin{tabular}{c|c|c|c|c}
disc. & $q=2$ & $q=4$ & $q=6$ & $q=8$ \\
\hline
1 & $\frac{3}{2} \pi < \omega < 1.85242 \pi$ & $\frac{3}{2} \pi < \omega \leq 1.92160 \pi$ & $\frac{3}{2} \pi < \omega \leq 1.92200 \pi$ & $\frac{3}{2} \pi < \omega < 1.92202 \pi$ \\
\hline
2 & $1.85242 \pi \leq \omega < 1.97057 \pi$ & $1.92160 \pi < \omega \leq 1.93839 \pi$ & $1.92200 \pi < \omega < 1.96903 \pi$ & $1.92202 \pi \leq \omega < 1.98410 \pi$ \\
\hline
3 & $1.93040 \pi \leq \omega \leq 1.97114 \pi$ & $1.92160 \pi < \omega \leq 1.99451 \pi$ & $1.92200 \pi < \omega \leq 1.99774 \pi$ & $1.92202 \pi \leq \omega < 1.99878 \pi$ \\
\hline
4 & & $1.94400 \pi \leq \omega \leq 1.99451 \pi$ & $1.94661 \pi < \omega \leq 1.99072 \pi$ & $1.94723 \pi \leq \omega < 1.99536 \pi$ \\
\hline
5 & & & $1.99072 \pi < \omega < 2\pi$ & $1.99536 \pi \leq \omega < 2\pi$ \\
\hline
6 & & & $1.99072 \pi < \omega \leq 1.99774 \pi$ & $1.99536 \pi \leq \omega < 1.99878 \pi$ \\
\end{tabular}
\label{table:triangularevenqdiscontinuities}
\end{center}
\end{minipage}
\end{table}

\begin{table}[p]
\begin{center}
\caption{Exchange interaction of even order $q$ in Hamiltonian (\ref{eqn:model}) and corresponding ranges of $\omega$ for the limits of the magnetization plateaus for the triangular lattice.}
\begin{tabular}{c|c|c}
$q$ & plateau 1 & plateau 2 \\
\hline
2 & $1.85242 \pi \leq \omega < 2\pi$ & $1.97114 \pi < \omega < 2\pi$ \\
\hline
4 & $1.92249 \pi \leq \omega < 2\pi$ & $1.99451 \pi < \omega < 2\pi$ \\
\hline
6 & $1.96776 \pi \leq \omega < 2\pi$ & $1.99774 \pi < \omega < 2\pi$ \\
\hline
8 & $1.98410 \pi \leq \omega < 2\pi$ & $1.99878 \pi \leq \omega < 2\pi$ \\
\end{tabular}
\label{table:triangularevenqplateaus}
\end{center}
\end{table}

\newpage

\begin{figure}
\begin{minipage}{\textwidth}
\centerline{
\includegraphics[width=0.3\textwidth,height=1.5in]{triangleexchangeq=5width.eps}
\hspace{0pt}
\includegraphics[width=0.3\textwidth,height=1.5in]{triangleexchangeq=7width.eps}
}
\vspace{17pt}
\centerline{
\includegraphics[width=0.3\textwidth,height=1.5in]{triangleexchangeq=4width.eps}
\hspace{0pt}
\includegraphics[width=0.3\textwidth,height=1.5in]{triangleexchangeq=6width.eps}
}
\caption{Magnetization change per spin $\frac{\Delta M}{N}$ for the discontinuities in the lowest energy configuration of Hamiltonian (\ref{eqn:model}) for the triangular lattice as a function of $\omega$ for (a) $q=5$, (b) $q=7$, (c) $q=4$, and (d) $q=6$. The numbering of the discontinuities follows Figs \ref{fig:qexchangetriangular1}(a), \ref{fig:qexchangetriangular1}(b), \ref{fig:qexchangetriangular1}(c), and \ref{fig:qexchangetriangular1}(d).
}
\label{fig:qexchangetriangular3}
\end{minipage}
\end{figure}

\begin{figure}
\includegraphics[width=3.5in,height=2.5in]{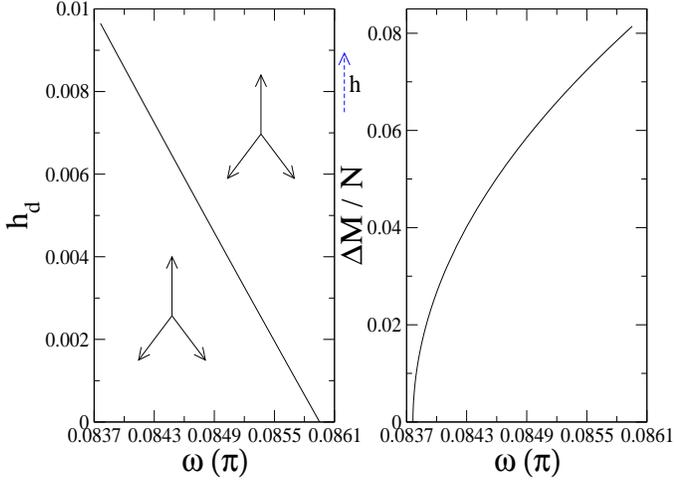}
\vspace{0pt}
\caption{(a) Magnetic field $h_d$ of discontinuity 4 (Fig. \ref{fig:qexchangetriangular1}(a)) in the lowest energy configuration of Hamiltonian (\ref{eqn:model}) for the triangular lattice as a function of $\omega$ for $q=5$. (b) Corresponding magnetization change per spin $\frac{\Delta M}{N}$ (Fig. \ref{fig:qexchangetriangular3}(a)).
}
\label{fig:qexchangetriangular4}
\end{figure}

\end{appendix}

\end{document}